\documentclass[12pt,tightenlines,eqsecnum,floats,aps,amsmath,amssymb,nofootinbib,prd,showpacs]{revtex4}

\usepackage{setspace} \usepackage{amsmath,amssymb,amsfonts,amsthm}
\usepackage{graphicx}

\newcommand{\be}{\begin{equation}}
\newcommand{\ee}{\end{equation}}
\newcommand{\beq}{\begin{eqnarray}}
\newcommand{\eeq}{\end{eqnarray}}

\def\lsim{\hbox{ \raise.35ex\rlap{$<$}\lower.6ex\hbox{$\sim$}\ }}
\def\gsim{\hbox{ \raise.35ex\rlap{$>$}\lower.6ex\hbox{$\sim$}\ }} 

\usepackage{bbold}

\begin{document}
\title{Unique factor ordering in the continuum limit of LQC}
\author{William Nelson and Mairi Sakellariadou} 
\affiliation{King's College London, Department of Physics, Strand WC2R 
2LS, London, U.K.}

\begin{abstract}
\vspace{.2cm}
\noindent
We show that the factor ordering ambiguities associated with the loop
quantisation of the gravitational part of the cosmological Hamiltonian
constraint, disappear at the level of Wheeler-DeWitt equation only for
a particular choice of lattice refinement model, which coincides with
constraints imposed from phenomenological and consistency arguments.
\end{abstract}
 

\maketitle

\section{Introduction}
Loop Quantum Gravity (LQG)~\cite{rovelli2004}, a non-perturbative and
background independent canonical quantisation of general relativity in
four space-time dimensions, is one of the main approaches to
quantising gravity. Even though the full theory of LQG is not yet
complete, its successes lead us to apply LQG techniques in a simple
setting where symmetry principles can be used. The application of LQG
to the cosmological sector, known as Loop Quantum Cosmology
(LQC)~\cite{Ashtekar:2003hd,Bojowald:2002gz}, has recently made
significant progress. The difference between LQC and other approaches
of quantum cosmology, is that the input is motivated by a full quantum
gravity theory. The simplicity of the setting (typically homogeneity and
then sometimes also isotropy, although recent progress has been made
towards inhomogeneous cosmologies~\cite{Rovelli:2008dx}), combined with the
discreteness of spatial geometry provided by LQG, render feasible the
overall study of LQC dynamics.

Loop quantum cosmology is formulated in terms of SU(2) holonomies of
the connection and triads. In LQC, the quantum evolution is described
by a second order difference equation, instead of the second order
differential equation of the Wheeler-DeWitt (WDW) approach to quantum
cosmology. As the scale factor increases, the universe eventually 
enters the semi-classical regime, and the WDW differential
equation describes, to a very good approximation, the subsequent
evolution.

In the 'old' quantisation, the quantised holonomies were taken to be
shift operators {\it with a fixed magnitude}, but later it was
found~\cite{Rosen:2006bga,Bojowald:2007ra} that this leads to
problematic instabilities in the continuum semi-classical limit.
Indeed, as the universe expands, the Hamiltonian constraint operator
creates new vertices of a lattice state, leading in LQC to a
refinement of the discrete lattice.  The effect of the lattice
refinement has been modelled and the elimination of the instabilities
in the continuum era has been explicitly shown~\cite{Nelson:2007um}.
Lattice refinement leads to new dynamical difference equations
which, in general, do not have a uniform step-size. Thus, their study
gets quite involved, particularly in the anisotropic cases. Recently,
numerical techniques have been developed~\cite{Sabharwal:2007xy, Nelson:2008}
to address this issue.

The {\sl correct} refinement model should be given by the full LQG
theory. One would have, in principle, to use the full Hamiltonian
constraint and find the way that its action balances the creation of
new vertices as the volume increases. Instead, phenomenological arguments have
been used, where the choice of the lattice refinement is constrained
by the form of the matter Hamiltonian~\cite{Nelson:2007um}. Here, we use
another argument to specif, within a particular class (power law), the
choice of the lattice refinement.

There are many equivalent ways of writing the Hamiltonian constraint
in terms of the triad and the holonomies of the connection, since at the
classical level holonomies commute. However, each of these factor
ordering choices leads to a different factor ordering of
the WDW equation in the continuum limit. In what follows we explicitly
demonstrate that the ambiguities at the classical limit of LQC, which
is precisely the WDW equation, disappear only for a particular choice
of a lattice refinement model.

\section{elements of LQC}

To quantise the gravitational Hamiltonian for isotropic flat
cosmologies, we restrict ourselves to an elementary (fiducial) cell
${\mathcal V}$, with finite fiducial volume; only in this volume
spatial integrations will be performed. This is the usual approach to
regularise divergences appearing in a quantisation scheme based on a
Hamiltonian framework within flat homogeneous models.

Introducing a flat fiducial metric $^0 q_{ab}$, in which the volume of
${\mathcal V}$ is ${\rm V}_0$, the phase space variables $p, c$ of
loop quantum cosmology read
\be |p| = {\rm V}_0^{3/2} \frac{a^2}{4} ~; \ \ \ \ c =
 \frac{{\rm V}_0^{1/3}}{2} \gamma \dot{a}~,
\ee
(with the lapse function set to 1) where $a$ is the 
cosmological scale factor and $\gamma $ is the
Barbero-Immirzi parameter, labelling in-equivalent quantum
theories. The triad component $p$, determining the physical volume of
the fiducial cell, is connected to the connection component $c$,
determining the physical edge length of the fiducial cell, through
the Poisson bracket
\be
\{c,p\}=  \frac{ \kappa\gamma}{3}~,
\ee
(with $\kappa=8\pi G$) which (for this choice of variables) is 
independent of the volume factor $V_0$.

The Hamiltonian formulation in the full LQG theory is based upon the
Ashtekar variables, namely the connection $A_a^{\rm i}$ and (density
weighted) triad $E^a_{\rm i}$, arising from a canonical transformation
of the ADM variables. Note that $i$ refers to the Lie algebra index 
and $a$ is a spatial index, with $a,i=1,2,3$. They are given by
\be
 A_a^{\rm i} = c{\rm V}_0^{-1/3}\   \omega^{\rm i}_a~; 
\ \ \ \ E^a_{\rm i} = p {\rm V}_0^{-2/3}
\sqrt{^0q}\  X^a_i~,
\ee
where $^0q$ is the determinant of the fiducial background metric
$^0q_{ab}=\omega^i_a\omega_{bi}$, with $\omega^i_a$ a basis of
  left-invariant one-forms, and $X^a_i$ are the Bianchi I basis vectors
  $X^a_i=\delta^a_i$.

After quantisation, states in the kinematical Hilbert space can be
expressed as (linear combinations of) eigenstates of $\hat{p}$, namely
\be 
\hat{p}
  |\mu\rangle = \frac{\kappa \gamma \hbar}{6} |\mu| |\mu\rangle~, 
\ee
which are diagonal, i.e. 
$\langle \mu_1 |\mu_2 \rangle =\delta_{\mu_1\mu_2}$. 

Just as in full LQG, there is no operator corresponding to the
connection, however the action of its holonomy is well defined,
\be
\label{eq:hol1} \hat{h}_{\rm
    i}|\mu\rangle = \left(\widehat{\rm Cs}{\mathbb 1} + 2\tau_{\rm i}
  \widehat{\rm Sn} \right)|\mu\rangle~, 
\ee 
where $\widehat{\rm Sn}$ and $\widehat{\rm Cs}$ are given by
\beq 
\widehat{\rm Sn}|\mu\rangle &=& \frac{1}{2i}\left(
e^{\frac{-i\widehat{\tilde{\mu}c}}{2}} -
e^{\frac{i\widehat{\tilde{\mu}c}}{2}} \right)|\mu\rangle =
\frac{-i}{2} \left( |\mu+\tilde{\mu} \rangle -
|\mu-\tilde{\mu}\rangle\right)~, \nonumber \\ \widehat{\rm
  Cs}|\mu\rangle &=& \frac{1}{2\ }\left(
e^{\frac{-i\widehat{\tilde{\mu}c}}{2}} +
e^{\frac{i\widehat{\tilde{\mu}c}}{2}} \right)|\mu\rangle =
\frac{\ \ 1}{\ 2} \left( |\mu+\tilde{\mu} \rangle +
|\mu-\tilde{\mu}\rangle\right)~. \label{eq:sin_cos}
\eeq 
When $\tilde{\mu}$ is a constant (typically called $\mu_0$ is the `old'
quantisation), it is clear that the operator $\exp 
\left(i\widehat{\tilde{\mu}c}/2\right)$ acts as a simple shift operator,
namely
\be 
\exp \left( i\widehat{\tilde{\mu}c}/2\right)|\mu\rangle =
  \exp\left(\widehat{\tilde{\mu}\frac{\rm d}{{\rm d}\mu} } \right)
  |\mu\rangle = |\mu+\tilde{\mu}\rangle~.  
\ee 
In the case of {\sl lattice refinement}, 
$\tilde{\mu}=\tilde{\mu}\left(\mu\right)$ is not a constant and
the shift interpretation is no longer valid. However, one can change
variables from $\mu$ to $\nu$:
\be
\label{eq:nu} 
\nu = k \int \frac{{\rm d} \mu}{\tilde{\mu}\left(\mu\right)}~, \ee 
for which 
\be 
e^{\widehat{\tilde{\mu}\frac{\rm d}{{\rm d}\mu}}} =
e^{\widehat{k \frac{\rm d}{{\rm d} \nu}}}~,
\ee 
with $k$ a constant. In these new variables the holonomies do act
as simple shift operators, with parameter length $k$, for states
labelled by $\nu$, defined as eigenvalues of 
$f\left(\hat{p}\right)$ (with $f$ the implicit function giving 
$\mu\left(\nu\right)$; it is obtained by solving Eq.~(\ref{eq:nu})).
Thus,
\be 
e^{\widehat{\tilde{\mu}\frac{\rm d}{{\rm d}\mu}}}|\nu\rangle =
e^{\widehat{k \frac{\rm d}{{\rm d} \nu}}}|\nu\rangle = |\nu+k\rangle~.
\ee 
One should keep in mind, that the relationship between $\nu$ and
geometric quantities, such as volume, is more
complicated than their relationship with $\mu$. In what follows, we
consider $\tilde{\mu}$ to be of
the form $\tilde{\mu} = \mu_0\mu^{-A}$, where $\mu_0$ is some
constant~\cite{Nelson:2007wj}. In this case, one can explicitly solve
Eq.~(\ref{eq:nu}) to obtain
\be
\label{eq:nu2} 
\nu = \frac{k\mu^{1-A}}{\mu_0 \left(1-A\right)}~.
\ee 
This one-parameter family of lattice refinement models includes the
`old' ($A=0$) and `new' ($A=-1/2$) quantisations. Motivated by the
full LQG theory, $A$ is expected to lie in the range
between $-1/2<A<0$~\cite{Bojowald:2007ra}. There are several phenomenological 
and consistency arguments supporting $A=-1/2$~\cite{Nelson:2007wj,Gorichi}, 
nevertheless here we keep $A$ as an 
undetermined parameter and show that the choice $A=-1/2$ leads to important 
consequences for the factor ordering of the continuum limit of the theory.

\section{Factor ordering} 

The classical gravitational part of the Hamiltonian constraint (with
the lapse function set to unity) is given by
\be
 {\mathcal C}_{\rm grav} = - \frac{1}{\gamma^2} \int _{\mathcal V}
 {\rm d}^3 x\ \epsilon _{\rm ijk} \frac{ E^{a{\rm i}}E^{b{\rm
       j}}F^{\rm k}_{ab}}{\sqrt{|\det E|}}~,
\ee
where $F^{\rm i}_{ab} = \partial _a A_b^{\rm i} - \partial_b A^{\rm
  i}_a + \epsilon^{\rm i}_{\rm jk} A_a^{\rm j}A_b^{\rm k}$ is the
curvature of the connection.  Writing this in terms of our quantisable
variables ($p$ and the holonomies of $c$), it reads~\cite{Ashtekar:2006wn}
\be\label{eq:ham}
\hat{\mathcal C}_{\rm grav} = \frac{2i \  
}{\kappa^2 \hbar \gamma^3 k^3}
 {\rm tr} \sum_{\rm ijk} \epsilon^{\rm ijk} 
\left( \hat{h}_{\rm i} \hat{h}_{\rm j}
\hat{h}_{\rm i}^{-1} \hat{h}_{\rm j}^{-1}\hat{h}_{\rm k} \left[ 
\hat{h}_{\rm k}^{-1},\hat{V} \right] \right)~.
\ee
There are many possible choices of factor ordering that could have
been made at this point, since classically the action of the
holonomies commute. Each of these possible choices will lead, in
principle, to a different factor ordering of the resulting continuum
WDW equation. 

We consider only factor orderings of the form of cyclic permutations
of the holonomy and volume operators within the trace. They have the
advantage of being trivially equivalent with respect to the spin
indices, while they remain within the irreducible representations of
the holonomies, hence avoiding the spurious, ill-behaved, solutions,
present for higher representations~\cite{Vandersloot:2005kh}. Whilst
these holonomies do not represent a complete set of factor ordering
choices, nevertheless they include the choices commonly made in the
literature~\cite{Ashtekar:2006wn,Vandersloot:2005kh}. Finally, using these factor
orderings we can demonstrate how different lattice refinement models
alter the factor ordering of the resulting WDW equation in the
continuum limit.

Using Eqs.~(\ref{eq:hol1}),(\ref{eq:sin_cos}), the action of the
different factor ordering choices give us explicitly:
\beq \label{eq:factor1}
\epsilon_{\rm ijk} {\rm tr}  \left( \hat{h}_{\rm i} \hat{h}_{\rm j}
\hat{h}_{\rm i}^{-1} \hat{h}_{\rm j}^{-1}\hat{h}_{\rm k} \left[ 
\hat{h}_{\rm k}^{-1},\hat{V} \right] \right) & = & 
-24\widehat{\rm Sn}^2\widehat{\rm Cs}^2 \left( \widehat{\rm Cs} \hat{V}
\widehat{\rm Sn} - \widehat{\rm Sn} \hat{V} \widehat{\rm Cs}\right)~, \\
\label{eq:factor2}
\epsilon_{\rm ijk} {\rm tr}  \left( \hat{h}_{\rm j}
\hat{h}_{\rm i}^{-1} \hat{h}_{\rm j}^{-1}\hat{h}_{\rm k} \left[ 
\hat{h}_{\rm k}^{-1},\hat{V} \right] \hat{h}_{\rm i}  \right) & = & 
-24\widehat{\rm Sn}^2\widehat{\rm Cs} \ \left( \widehat{\rm Cs} \hat{V}
\widehat{\rm Sn} - \widehat{\rm Sn} \hat{V} \widehat{\rm Cs}\right) \widehat{\rm Cs}~, \\
\label{eq:factor3}
\epsilon_{\rm ijk} {\rm tr}  \left( 
\hat{h}_{\rm i}^{-1} \hat{h}_{\rm j}^{-1}\hat{h}_{\rm k} \left[ 
\hat{h}_{\rm k}^{-1},\hat{V} \right] \hat{h}_{\rm i}  \hat{h}_{\rm j} \right) & = & 
 \ \ \ \ -12\widehat{\rm Sn}^2 \left( \widehat{\rm Cs} \hat{V}
\widehat{\rm Sn} - \widehat{\rm Sn} \hat{V} \widehat{\rm Cs} \right)\widehat{\rm Cs}^2 \nonumber \\
&& \ \ \ \ -12\widehat{\rm Cs}^2 \left( \widehat{\rm Cs} \hat{V}
\widehat{\rm Sn} - \widehat{\rm Sn} \hat{V} \widehat{\rm Cs} \right)\widehat{\rm Sn}^2~, \\
\label{eq:factor4}
\epsilon_{\rm ijk} {\rm tr}  \left( 
\hat{h}_{\rm j}^{-1}\hat{h}_{\rm k} \left[ 
\hat{h}_{\rm k}^{-1},\hat{V} \right] \hat{h}_{\rm i} \hat{h}_{\rm j} \hat{h}_{\rm i}^{-1} \right) & = & 
 \ \ \ \ -24\widehat{\rm Cs} \ \left( \widehat{\rm Cs} \hat{V}
\widehat{\rm Sn} - \widehat{\rm Sn} \hat{V} \widehat{\rm Cs}\right) \widehat{\rm Sn}^2 \widehat{\rm Cs}~, \\
\label{eq:factor5}
\epsilon_{\rm ijk} {\rm tr}  \left( 
\hat{h}_{\rm k} \left[ 
\hat{h}_{\rm k}^{-1},\hat{V} \right] \hat{h}_{\rm i} \hat{h}_{\rm j} \hat{h}_{\rm i}^{-1} \hat{h}_{\rm j}^{-1}
\right) & = & 
 \ \ \ \ \ \ \ \ -24\left( \widehat{\rm Cs} \hat{V}
\widehat{\rm Sn} - \widehat{\rm Sn} \hat{V} \widehat{\rm Cs}\right) \widehat{\rm Sn}^2 \widehat{\rm Cs}^2~, \\
\label{eq:factor6}
\epsilon_{\rm ijk} {\rm tr}  \left( 
 \left[ 
\hat{h}_{\rm k}^{-1},\hat{V} \right] \hat{h}_{\rm i} \hat{h}_{\rm j} \hat{h}_{\rm i}^{-1} \hat{h}_{\rm j}^{-1}
\hat{h}_{\rm k} \right) & = & 
 \ \ \ \ \ \ \ \ -24\left( \widehat{\rm Cs} \hat{V}
\widehat{\rm Sn} - \widehat{\rm Sn} \hat{V} \widehat{\rm Cs}\right) \widehat{\rm Sn}^2 \widehat{\rm Cs}^2~,
\eeq
where we have made extensive use of the trace identities:
\beq
 {\rm tr}\left( \tau_{\rm i}\right) &=& \frac{-i}{2} {\rm tr} \left( \sigma_{\rm i} \right) = 0 \nonumber \\
 {\rm tr}\left( \tau_{\rm i}\tau_{\rm j}\right) &=& \frac{-1}{4} {\rm tr} \left( \sigma_{\rm i} \sigma_{\rm j}
 \right) = \frac{-1}{2}\delta_{\rm ij} \nonumber \\
 {\rm tr}\left( \tau_{\rm i}\tau_{\rm j}\tau_{\rm k}\right) &=& \frac{i}{8} {\rm tr} \left( \sigma_{\rm i} \sigma_{\rm j}
\sigma_{\rm k} \right) = \frac{-1}{4}\epsilon_{\rm jki} \nonumber \\
 {\rm tr}\left(\tau_{\rm j} \tau_{\rm i}\tau_{\rm j}\tau_{\rm k}\right) &=& \frac{1}{16} {\rm tr} \left( \sigma_{\rm j}
\sigma_{\rm i} \sigma_{\rm j} \sigma_{\rm k} \right) = \frac{-1}{8}\delta_{\rm ik} + \frac{1}{4}\epsilon_{\rm ijl}\epsilon_{\rm
kjl}~.
\eeq
Using Eqs.~(\ref{eq:sin_cos}) and defining $\hat{V}|\nu\rangle =
V_{\nu}|\nu\rangle$, the action of each of these factor orderings on
the basis $|\nu\rangle$ can be calculated.  For clarity,
we will derive the continuum limit only for the factor ordering choice
given in Eq.~(\ref{eq:factor1}); the other ones follow along similar lines 
and for completeness have been included in an appendix (Appendix A).

Thus,
\be
\widehat{\rm Sn}^2\widehat{\rm Cs}^2 \left( \widehat{\rm Cs} V
\widehat{\rm Sn} - \widehat{\rm Sn} V \widehat{\rm
  Cs}\right)|\nu\rangle = \frac{-i}{32} \left( V_{\nu+k} -V_{\nu-k}
\right)\Bigl( |\nu+4k\rangle -2|\nu\rangle +|\nu-4k\rangle \Bigr)~.
\ee
Extending the above, one can obtain the action of the chosen factor ordering 
on a general state in the Hilbert space given by 
$|\Psi \rangle = \sum_\nu \psi_\nu|\nu\rangle$. The 
explicit difference equations for the coefficients $\psi_\nu$ read
\beq\label{eq:ham1.2}
\epsilon_{\rm ijk} {\rm tr}  \left( \hat{h}_{\rm i} \hat{h}_{\rm j}
\hat{h}_{\rm i}^{-1} \hat{h}_{\rm j}^{-1}\hat{h}_{\rm k} \left[ 
\hat{h}_{\rm k}^{-1},\hat{V} \right] \right)|\Psi\rangle & = & 
\frac{-3i}{4} \sum_\nu \Biggl[ \Bigl( V_{\nu-3k} - V_{\nu-5k} \Bigr) 
\psi_{\nu-4k}
-2\Bigl(V_{\nu+k}- V_{\nu-k}\Bigr)\psi_\nu \Biggr. \nonumber \\
&&+\Biggl. \Bigl( V_{\nu+5k} - V_{\nu+3k} \Bigr) \psi_{\nu+4k} \Biggr] |
\nu\rangle~.
\eeq
We can now take the continuum limit of these expressions by expanding
$\psi_\nu \approx \psi\left(\nu\right)$ as a Taylor expansion in small
$k/\nu$, i.e. in the limit that the discreteness scale ($k$) is much
smaller than the scale of the universe (which is given by $\nu$). By
noting that the volume is given by
\be
 V_{\nu}|\nu\rangle \sim [\mu\left(\nu\right)]^{3/2}|\nu\rangle~,
\ee
where $\mu\left(\nu\right)$ is obtained by Eq.~(\ref{eq:nu2}) and we are
neglecting a constant factor $\left( \kappa \gamma
\hbar/6\right)^{3/2}$, we find 
\be
V_{\nu\pm nk} \sim \Bigl[ \left( \nu \pm nk\right) 
\alpha \Bigr]^{3/\left[2(1-A)\right]}~,
\ee
where $\alpha = \mu_0\left(1-A\right)/k$. 

In general, the above needs also to be expanded in the $k/\nu
\rightarrow 0$ limit and due to the $k^{-3}$ factor in
Eq.~(\ref{eq:ham}) it is necessary to go to third order in both this and
the Taylor expansion. To expand Eq.~(\ref{eq:ham1.2}) we need the
difference between the volume eigenvalues evaluated on different
lattice points, given by
\beq\label{eq:vol}
V_{\nu+nk} - V_{\nu+mk} &\sim& \frac{3k}{2\left(1-A\right)}\alpha^{3/\left[2\left(1-A\right)\right]} \nu^{\left(1+2A\right)/
\left[2\left(1-A\right)\right]}
\Biggl[ \left( n-m\right) + \frac{1+2A}{4\left(1-A\right)} \frac{k}{\nu} \left(n^2-m^2\right) \Biggr.\nonumber \\ 
&&\Biggl.+ \frac{\left( 1+2A\right)\left(4A-1\right)}{24\left(1-A\right)^2} \frac{k^2}{\nu^2} \left(
n^3 - m^3\right) +{\cal O}\left( \frac{k^3}{\nu^3}\right) \Biggr]~.
\eeq
Performing a Taylor expansion of Eq.~(\ref{eq:ham1.2}) we get
the large scale continuum limit of the Hamiltonian constraint:
\beq\label{eq:final}
\lim_{k/\nu \rightarrow 0} 
\epsilon_{\rm ijk} {\rm tr}  \left( \hat{h}_{\rm i} \hat{h}_{\rm j}
\hat{h}_{\rm i}^{-1} \hat{h}_{\rm j}^{-1}\hat{h}_{\rm k} \left[ 
\hat{h}_{\rm k}^{-1},\hat{V} \right] \right)|\Psi\rangle \sim  
\nonumber \\
\frac{-36i}{1-A} \alpha^{3/\left[2\left(1-A\right)\right]} 
k^3 \sum_\nu \nu^{\left(1+2A\right)/\left[2\left(
1-A\right)\right]}
 \Biggl[ \frac{{\rm d}^2 \psi}{{\rm d} \nu^2 } + 
\frac{1+2A}{1-A} \frac{1}{\nu} \frac{{\rm d}
\psi}{{\rm d}\nu} + \frac{\left(1+2A\right)
\left(4A-1\right)}{\left(1-A\right)^2} \frac{1}{4\nu^2}
\psi\left(\nu\right)
\Biggr] |\nu\rangle~, \nonumber \\ 
\eeq
Taking $A=0$ reproduces the large scale factor ordering associated
with the `old' quantisation, as expected~\cite{Nelson:2007um}.  Notice that 
$A=-1/2$, which corresponds to the `new' quantisation, leads to the
following very simple form of the evolution equation in the 
continuum limit:
\be
\lim_{k/\nu \rightarrow 0} {\cal C}_{\rm grav} |\Psi\rangle = 
\frac{72 \mu_0/k}{\kappa^2 \hbar \gamma^3}
\left(\frac{\kappa\gamma\hbar}{6}\right)^{3/2} \sum_\nu \frac{{\rm d}^2
\psi}{{\rm d} \nu^2} |\nu\rangle~,
\ee
where we have used $\alpha = 3\mu_0/(2k)$ and reintroduced all the
constants. 

For $\mu_0 = k$ we arrive at the following final result
for the continuum limit of the WDW equation:
\be\label{eq:final_con}
\lim_{k/\nu \rightarrow 0} {\cal C}_{\rm grav} |\Psi\rangle = 
\frac{72}{\kappa^2 \hbar \gamma^3}
\left(\frac{\kappa\gamma\hbar}{6}\right)^{3/2} \sum_\nu \frac{{\rm d}^2
\psi}{{\rm d} \nu^2} |\nu\rangle~.
\ee

Repeating this tedious, but straight forward calculation for the other
factor ordering choices given in
Eqs.~(\ref{eq:factor2})-(\ref{eq:factor6}) results in the following
differential equations in the continuum limit (see the Appendix A
for details),
\beq
\lim_{k/\nu \rightarrow 0} 
\epsilon_{\rm ijk} {\rm tr}  \left( \hat{h}_{\rm j}
\hat{h}_{\rm i}^{-1} \hat{h}_{\rm j}^{-1}\hat{h}_{\rm k} \left[ 
\hat{h}_{\rm k}^{-1},\hat{V} \right] \hat{h}_{\rm i} \right)|
\Psi\rangle  \sim && \nonumber \\
\frac{-36i}{1-A} \alpha^{3/\left[2\left(1-A\right)\right]} \nonumber\\
\times k^3 \sum_\nu \nu^{\left(1+2A\right)/\left[2\left(
1-A\right)\right]}
 \Biggl[ \frac{{\rm d}^2 \psi}{{\rm d} \nu^2 } + 
\frac{1+2A}{1-A} \frac{1}{\nu} \frac{{\rm d}
\psi}{{\rm d}\nu} + \frac{\left(1+2A\right)
\left(4A-1\right)}{\left(1-A\right)^2} \frac{1}{4\nu^2}
\psi\left(\nu\right)
\Biggr] |\nu\rangle~, 
\eeq

\beq
\lim_{k/\nu \rightarrow 0} 
\epsilon_{\rm ijk} {\rm tr}  \left( 
\hat{h}_{\rm i}^{-1} \hat{h}_{\rm j}^{-1}\hat{h}_{\rm k} \left[ 
\hat{h}_{\rm k}^{-1},\hat{V} \right]
\hat{h}_{\rm i} \hat{h}_{\rm j} \right)|\Psi\rangle \sim \nonumber \\
\frac{-36i}{1-A} \alpha^{3/\left[2\left(1-A\right)\right]}\nonumber\\
\times k^3 \sum_\nu \nu^{\left(1+2A\right)/\left[2\left(
1-A\right)\right]}
 \Biggl[ \frac{{\rm d}^2 \psi}{{\rm d} \nu^2 } + \frac{1+2A}{1-A} 
\frac{1}{2\nu} \frac{{\rm d}
\psi}{{\rm d}\nu} 
+ \frac{\left(1+2A\right)\left(4A-1\right)}{\left(1-A\right)^2} 
\frac{1}{8\nu^2}\psi\left(\nu\right)
\Biggr] |\nu\rangle~, 
\eeq

\beq
\lim_{k/\nu \rightarrow 0} 
\epsilon_{\rm ijk} {\rm tr}  \left( 
\hat{h}_{\rm j}^{-1}\hat{h}_{\rm k} \left[ 
\hat{h}_{\rm k}^{-1},\hat{V} \right]
\hat{h}_{\rm i} \hat{h}_{\rm j} \hat{h}_{\rm i}^{-1} \right)
|\Psi\rangle =\nonumber\\
\lim_{k/\nu \rightarrow 0} 
\epsilon_{\rm ijk} {\rm tr}  \left( 
\hat{h}_{\rm k} \left[ 
\hat{h}_{\rm k}^{-1},\hat{V} \right]
\hat{h}_{\rm i} \hat{h}_{\rm j}
 \hat{h}_{\rm i}^{-1} \hat{h}_{\rm j}^{-1} \right)|\Psi\rangle = \nonumber \\
\lim_{k/\nu \rightarrow 0} 
\epsilon_{\rm ijk} {\rm tr}  \left( \left[ 
\hat{h}_{\rm k}^{-1},\hat{V} \right]
\hat{h}_{\rm i} \hat{h}_{\rm j}
 \hat{h}_{\rm i}^{-1} \hat{h}_{\rm j}^{-1} \hat{h}_{\rm k} \right)|\Psi\rangle 
\sim \nonumber\\
\frac{-36i}{1-A} \alpha^{3/\left(2\left(1-A\right)\right)} k^3 
\sum_\nu \nu^{\left(1+2A\right)/\left(2\left(
1-A\right)\right)}
 \frac{{\rm d}^2 \psi}{{\rm d} \nu^2 } |\nu\rangle~. \nonumber \\
\eeq
Once again we see that $A=-1/2$ results in a particular
simplification, in that \emph{all} of the considered factor ordering
choices reduce to Eq.~(\ref{eq:final_con}). This is to be expected
since quantum factor ordering ambiguities should disappear at the
classical level. The crucial finding is that the LQC ambiguities,
associated with the factor ordering in the Hamiltonian constraint
disappear at the continuum described by the WDW equation, {\it only} for a
lattice refinement power law model with $A=-1/2$. Thus, only this
model has a non-ambiguous continuum limit.

In addition to this lattice refinement model providing a unique choice
of factor ordering for the continuum limit equations, the action of
the volume operator is greatly simplified,
\be
 V_{\nu+nk} - V_{\nu +mk} = \left(n-m\right)\frac{3\mu_0}{2}~.
\ee
This is no accident since the initial motivation for this quantisation
procedure was that the volume, rather than the area, should
get quantised. The consequence is that quantum corrections to this
classical equation (i.e. quantum corrections to general relativity)
enter only in the Taylor approximation $\psi(\nu+nk) \approx \psi(\nu)
+ \dots$ and not in the expansion of the volume terms,
i.e. Eq.~(\ref{eq:vol}) requires no approximation.

Finally, to relate this result to more usual variables, we can use
$\mu \sim p = a^2$ and $\nu \sim \mu^{3/2}$, to find that the factor
ordering of the Wheeler-DeWitt equation predicted by the large scale
limit of LQC reads
\be
 {\mathcal C}_{\rm grav} \sim \frac{{\rm d}^2 \psi}{{\rm d} \nu^2} 
\sim a^{-2} \frac{\rm d}{{\rm d}a}
\left( a^{-2} \frac{{\rm d} \psi}{{\rm d}a} \right)~,
\ee
where constants, but no factors on $a$, have been dropped.

\section{Conclusions}
Just as in the quantisation of standard fields, loop quantum cosmology
results in factor ordering ambiguities. One expects that the classical
limit should unambiguously result to the original classical
equation. Here, we have shown that in general this is not true at the
level of the Wheeler-DeWitt equation, which is the classical limit of
LQC (it is obtained when the discreteness scale is set to zero).  This
ordering ambiguity disappears however for the particular lattice
refinement model given by $A=-1/2$, which is typically called `new' or
`improved' quantisation.

The work presented here can be viewed in two ways: One could accept
the phenomenological~\cite{Nelson:2007wj,Nelson:2007um} and
consistency~\cite{Gorichi} requirements indicating that $A=-1/2$ is
the correct quantisation approach. In this way, we have shown that LQC
predicts a unique factor ordering of the Wheeler-DeWitt equation in
its continuum limit (at least for the particular class of factor
ordering considered here).  Alternatively, one could require that
factor ordering ambiguities in LQC should disappear at the level of
the Wheeler-DeWitt equation. In this case, we have shown that the
lattice refinement model should be $A=-1/2$.  In either case, we have
clearly demonstrated that there is a strong link between factor
ordering and lattice refinement in LQC; these two ambiguities are
closely related. In fact, we have shown that by specifying a
particular lattice refinement model we can uniquely determine the
factor ordering of the equation in the continuum limit, and vice
versa.

In conclusion, it is remarkable that the requirement for the
Wheeler-DeWitt factor ordering to be unique, is precisely the same
requirement reached by physical considerations of large scale physics
and consistency of the quantisation structure. In particular, it has
been previously shown~\cite{Nelson:2007wj} that for LQC to generically
support inflation and other matter fields without the onset of large
scale quantum gravity corrections, $A$ should be equal to $-1/2$. It
has been recently shown~\cite{Gorichi} that physical quantities depend
on the choice of the elementary cell used to regulate the spatial
integrations, unless one chooses $A=-1/2$, and that sensible effective
equations exist only for this choice. Taking this together with the
uniqueness of the factor ordering of the Wheeler-DeWitt equation, which
as we have shown also requires $A=-1/2$, it is clear that several
vastly different unrelated approaches have converged on the same
restriction of the theory. It is possible that this restriction can be
used to improve our understanding of the underlying full loop quantum
gravity theory.

\vskip.05truecm
\acknowledgments This work is partially supported by the European
Union through the Marie Curie Research and Training Network
\emph{UniverseNet} (MRTN-CT-2006-035863).

\appendix
\label{appendix}
\section{}
The action of the different factor ordering possibilities considered in 
Eqs.~(\ref{eq:factor2})-(\ref{eq:factor6})
on a basis state $|\nu\rangle$ is given by
\beq
\widehat{\rm Sn}^2\widehat{\rm Cs} \ \left( \widehat{\rm Cs} \hat{V}
\widehat{\rm Sn} - \widehat{\rm Sn} \hat{V} \widehat{\rm Cs}\right) \widehat{\rm Cs} |\nu\rangle&=&
\frac{-i}{32} \Biggl[ \left( V_{\nu+2k} - V_{\nu} \right) \Bigl( |\nu+4k\rangle-|\nu+2k\rangle
-|\nu\rangle +|\nu-2k\rangle \Bigr) \Biggr. \nonumber \\
&&\Biggl. +\left( V_{\nu} - V_{\nu-2k} \right) \Bigl( |\nu+2k\rangle-|\nu\rangle
-|\nu-2k\rangle +|\nu-4k\rangle \Bigr) \Biggr] \nonumber \\ \\
\widehat{\rm Sn}^2 \left( \widehat{\rm Cs} \hat{V}
\widehat{\rm Sn} - \widehat{\rm Sn} \hat{V} \widehat{\rm Cs} \right)\widehat{\rm Cs}^2|\nu\rangle &=&
\frac{-i}{32} \Biggl[ \left( V_{\nu +3k} - V_{\nu+k}\right)\Bigl( |\nu +4k\rangle - 2|\nu+2k\rangle
+|\nu\rangle \Bigr) \Biggr. \nonumber \\
&&\Biggl.+ 2\left(V_{\nu+k}-V_{\nu-k}\right) \Bigl( |\nu+2k\rangle -2|\nu\rangle+|\nu-2k\rangle \Bigr) \Biggr. \nonumber \\
&&\Biggl.+ \left( V_{\nu-k} - V_{\nu -3k} \right) \Bigl( |\nu\rangle -2|\nu-2k\rangle +|\nu-4k\rangle \Bigr) \Biggr]~, \\
\widehat{\rm Cs}^2 \left( \widehat{\rm Cs} \hat{V}
\widehat{\rm Sn} - \widehat{\rm Sn} \hat{V} \widehat{\rm Cs} \right)\widehat{\rm Sn}^2|\nu\rangle &=&
\frac{-i}{32} \Biggl[ \left( V_{\nu +3k} - V_{\nu+k}\right)\Bigl( |\nu +4k\rangle + 2|\nu+2k\rangle
+|\nu\rangle \Bigr) \Biggr. \nonumber \\
&&\Biggl.- 2\left(V_{\nu+k}-V_{\nu-k}\right) \Bigl( |\nu+2k\rangle +2|\nu\rangle+|\nu-2k\rangle \Bigr) \Biggr. \nonumber \\
&&\Biggl.+ \left( V_{\nu-k} - V_{\nu -3k} \right) \Bigl( |\nu\rangle +2|\nu-2k\rangle +|\nu-4k\rangle \Bigr) \Biggr]~,
\eeq
\beq
\widehat{\rm Cs} \left( \widehat{\rm Cs} \hat{V}
\widehat{\rm Sn} - \widehat{\rm Sn} \hat{V} \widehat{\rm Cs} \right)\widehat{\rm Sn}^2\widehat{\rm Cs}|\nu\rangle &=&
\frac{-i}{32} \Biggl[ \left( V_{\nu +4k} - V_{\nu+2k}\right)\Bigl( |\nu +4k\rangle + |\nu+2k\rangle \Bigr) \Biggr. \nonumber \\
&&\Biggl.- \left(V_{\nu+2k}-V_{\nu}\right) \Bigl( |\nu+2k\rangle +|\nu\rangle \Bigr) \Biggr. \nonumber \\
&&\Biggl.- \left(V_{\nu}-V_{\nu-2k}\right) \Bigl( |\nu\rangle +|\nu-2k\rangle \Bigr) \Biggr. \nonumber \\
&&\Biggl.+ \left( V_{\nu-2k} - V_{\nu -4k} \right) \Bigl( |\nu-2k\rangle +|\nu-4k\rangle \Bigr) \Biggr]~, \\
\left( \widehat{\rm Cs} \hat{V} \widehat{\rm Sn} - \widehat{\rm Sn} \hat{V} \widehat{\rm Cs} \right)
\widehat{\rm Sn}^2\widehat{\rm Cs}^2|\nu\rangle &=&
\frac{-i}{32} \Biggl[ \left( V_{\nu +5k} - V_{\nu+3k}\right)|\nu+4k\rangle
-2 \left(V_{\nu+k}-V_{\nu-k}\right) |\nu\rangle \Biggr. \nonumber \\
&&\Biggl.+ \left( V_{\nu-3k} - V_{\nu -5k} \right) |\nu-4k\rangle \Biggr]~.
\eeq
Extending these to a general state in the Hilbert space given by $|\Psi\rangle = \sum_\nu \psi_\nu|\nu\rangle$ gives,
\beq
\label{eq:ham2.2}
\epsilon_{\rm ijk} {\rm tr}  \left( \hat{h}_{\rm j}
\hat{h}_{\rm i}^{-1} \hat{h}_{\rm j}^{-1}\hat{h}_{\rm k} \left[ 
\hat{h}_{\rm k}^{-1},\hat{V} \right] \hat{h}_{\rm i}  \right)|\Psi\rangle & = & 
\frac{-3i}{4} \sum_\nu \Biggl( \Bigl( V_{\nu-2k} - V_{\nu-4k} \Bigr) \psi_{\nu-4k}
-\Bigl(V_{\nu}- V_{\nu-2k}\Bigr)\psi_{\nu-2k} \Biggr. \nonumber \\
&&- \Bigl( V_{\nu+2k} - V_{\nu} \Bigr) \psi_{\nu} + \Bigl( V_{\nu+4k} - V_{\nu+2k}\Bigr)\psi_{\nu+2k}
\nonumber \\
&&+ \Bigl( V_{\nu-2k} - V_{\nu-4k} \Bigr) \psi_{\nu-2k} - \Bigl( V_{\nu} - V_{\nu-2k}\Bigr)\psi_{\nu}
\nonumber \\
&& \Biggl. - \Bigl( V_{\nu+2k} - V_{\nu} \Bigr) \psi_{\nu+2k} + \Bigl( V_{\nu+4k} - V_{\nu+2k}\Bigr)\psi_{\nu+4k}
\Biggr) |\nu\rangle~, 
\eeq
%
\beq
\label{eq:ham3.2}
\epsilon_{\rm ijk} {\rm tr}  \left( 
\hat{h}_{\rm i}^{-1} \hat{h}_{\rm j}^{-1}\hat{h}_{\rm k} \left[ 
\hat{h}_{\rm k}^{-1},\hat{V} \right] \hat{h}_{\rm i}  \hat{h}_{\rm j} \right) |\Psi\rangle & = & 
\frac{-3i}{4} \sum_\nu \Biggl( \Bigl( V_{\nu-k} - V_{\nu-3k} \Bigr) \psi_{\nu-4k}
+\Bigl(V_{\nu+3k}- V_{\nu+k}\Bigr)\psi_{\nu} \Biggr. \nonumber \\
&&- 4\Bigl( V_{\nu+k} - V_{\nu-k} \Bigr) \psi_{\nu} + \Bigl( V_{\nu-k} - V_{\nu-3k}\Bigr)\psi_{\nu}
\nonumber \\
&&\Biggl. + \Bigl( V_{\nu+3k} - V_{\nu+k} \Bigr) \psi_{\nu+4k} \Biggr)|\nu\rangle~,
\eeq
%
\beq
\label{eq:ham4.2}
\epsilon_{\rm ijk} {\rm tr}  \left( 
\hat{h}_{\rm j}^{-1}\hat{h}_{\rm k} \left[ 
\hat{h}_{\rm k}^{-1},\hat{V} \right] \hat{h}_{\rm i} \hat{h}_{\rm j} \hat{h}_{\rm i}^{-1} \right)|\Psi\rangle & = & 
\frac{-3i}{4} \sum_\nu \Biggl( \Bigl( V_{\nu} - V_{\nu-2k} \Bigr) \psi_{\nu-4k}
+\Bigl(V_{\nu+2k}- V_{\nu}\Bigr)\psi_{\nu-2k} \Biggr. \nonumber \\
&&- \Bigl( V_{\nu} - V_{\nu-2k} \Bigr) \psi_{\nu-2k} - \Bigl( V_{\nu+2k} - V_{\nu}\Bigr)\psi_{\nu}
\nonumber \\
&&- \Bigl( V_{\nu} - V_{\nu-2k} \Bigr) \psi_{\nu} - \Bigl( V_{\nu+2k} - V_{\nu}\Bigr)\psi_{\nu+2k}
\nonumber \\
&&\Biggl. + \Bigl( V_{\nu} - V_{\nu-2k} \Bigr) \psi_{\nu+2k} + \Bigl( V_{\nu+2k} - V_{\nu} \Bigr) \psi_{\nu+4k}
\Biggr)|\nu\rangle~,
\eeq
%
\beq
\label{eq:ham5.2}
\epsilon_{\rm ijk} {\rm tr}  \left( 
\hat{h}_{\rm k} \left[ 
\hat{h}_{\rm k}^{-1},\hat{V} \right] \hat{h}_{\rm i} \hat{h}_{\rm j} \hat{h}_{\rm i}^{-1} \hat{h}_{\rm j}^{-1}
\right) |\Psi\rangle & = & \nonumber \\
\epsilon_{\rm ijk} {\rm tr}  \left( 
 \left[ 
\hat{h}_{\rm k}^{-1},\hat{V} \right] \hat{h}_{\rm i} \hat{h}_{\rm j} \hat{h}_{\rm i}^{-1} \hat{h}_{\rm j}^{-1}
\hat{h}_{\rm k} \right) |\Psi\rangle & = &\nonumber \\
\frac{-3i}{4} \sum_\nu \Biggl( \Bigl( V_{\nu+k} - V_{\nu-k} \Bigr) \Bigl(\psi_{\nu-4k}
-2\psi_{\nu}+  \psi_{\nu+4k} \Bigr) \Biggr)|\nu\rangle~.\nonumber \\
\eeq
Performing the Taylor expansion of these as in Eq.~(\ref{eq:final}) we get,

\beq
\lim_{k/\nu \rightarrow 0} 
\epsilon_{\rm ijk} {\rm tr}  \left( \hat{h}_{\rm j}
\hat{h}_{\rm i}^{-1} \hat{h}_{\rm j}^{-1}\hat{h}_{\rm k} \left[ 
\hat{h}_{\rm k}^{-1},\hat{V} \right] \hat{h}_{\rm i} \right)|\Psi\rangle  
\sim \nonumber \\
\frac{-36i}{1-A} \alpha^{3/\left(2\left(1-A\right)\right)}  \nonumber \\
\times k^3 \sum_\nu \nu^{\left(1+2A\right)/\left(2\left(
1-A\right)\right)}
 \Biggl( \frac{{\rm d}^2 \psi}{{\rm d} \nu^2 } + \frac{1+2A}{1-A} \frac{1}{\nu} \frac{{\rm d}
\psi}{{\rm d}\nu} + \frac{\left(1+2A\right)\left(4A-1\right)}{\left(1-A\right)^2} \frac{1}{4\nu^2}\psi\left(\nu\right)
\Biggr) |\nu\rangle~, 
\eeq

\beq
\lim_{k/\nu \rightarrow 0} 
\epsilon_{\rm ijk} {\rm tr}  \left( 
\hat{h}_{\rm i}^{-1} \hat{h}_{\rm j}^{-1}\hat{h}_{\rm k} \left[ 
\hat{h}_{\rm k}^{-1},\hat{V} \right]
\hat{h}_{\rm i} \hat{h}_{\rm j} \right)|\Psi\rangle \sim \nonumber \\
\frac{-36i}{1-A} \alpha^{3/\left(2\left(1-A\right)\right)} \nonumber \\
\times k^3 \sum_\nu \nu^{\left(1+2A\right)/\left(2\left(
1-A\right)\right)}
 \Biggl( \frac{{\rm d}^2 \psi}{{\rm d} \nu^2 } + \frac{1+2A}{1-A} \frac{1}{2\nu} \frac{{\rm d}
\psi}{{\rm d}\nu} + \frac{\left(1+2A\right)\left(4A-1\right)}{\left(1-A\right)^2} \frac{1}{8\nu^2}\psi\left(\nu\right)
\Biggr) |\nu\rangle~, 
\eeq
\nonumber \\ 
\beq
\lim_{k/\nu \rightarrow 0} 
\epsilon_{\rm ijk} {\rm tr}  \left( 
\hat{h}_{\rm j}^{-1}\hat{h}_{\rm k} \left[ 
\hat{h}_{\rm k}^{-1},\hat{V} \right]
\hat{h}_{\rm i} \hat{h}_{\rm j} \hat{h}_{\rm i}^{-1} \right)|\Psi\rangle = 
 \nonumber \\
\lim_{k/\nu \rightarrow 0} 
\epsilon_{\rm ijk} {\rm tr}  \left( 
\hat{h}_{\rm k} \left[ 
\hat{h}_{\rm k}^{-1},\hat{V} \right]
\hat{h}_{\rm i} \hat{h}_{\rm j}
 \hat{h}_{\rm i}^{-1} \hat{h}_{\rm j}^{-1} \right)|\Psi\rangle = \nonumber \\
\lim_{k/\nu \rightarrow 0} 
\epsilon_{\rm ijk} {\rm tr}  \left( \left[ 
\hat{h}_{\rm k}^{-1},\hat{V} \right]
\hat{h}_{\rm i} \hat{h}_{\rm j}
 \hat{h}_{\rm i}^{-1} \hat{h}_{\rm j}^{-1} \hat{h}_{\rm k} \right)|\Psi\rangle 
\sim\nonumber \\
\frac{-36i}{1-A} \alpha^{3/\left(2\left(1-A\right)\right)} 
k^3 \sum_\nu \nu^{\left(1+2A\right)/\left(2\left(
1-A\right)\right)}
 \frac{{\rm d}^2 \psi}{{\rm d} \nu^2 } |\nu\rangle~. \nonumber \\
\eeq


\end{document}